# Estimation of Simultaneously Sparse and Low Rank Matrices

**Emile Richard**                                EMILE.RICHARD@CMLA.ENS-CACHAN.FR
CMLA UMR CNRS 8536, ENS Cachan & 1000mercis, France

**Pierre-André Savalle**                         PIERRE-ANDRE.SAVALLE@ECP.FR
Ecole Centrale Paris, France

**Nicolas Vayatis**                              NICOLAS.VAYATIS@CMLA.ENS-CACHAN.FR
CMLA UMR CNRS 8536, ENS Cachan, France

## Abstract

The paper introduces a penalized matrix estimation procedure aiming at solutions which are sparse and low-rank at the same time. Such structures arise in the context of social networks or protein interactions where underlying graphs have adjacency matrices which are block-diagonal in the appropriate basis. We introduce a convex mixed penalty which involves $\ell_1$-norm and trace norm simultaneously. We obtain an oracle inequality which indicates how the two effects interact according to the nature of the target matrix. We bound generalization error in the link prediction problem. We also develop proximal descent strategies to solve the optimization problem efficiently and evaluate performance on synthetic and real data sets.

## 1. Introduction

Matrix estimation is at the center of many modern applications and theoretical advances in the field of high dimensional statistics. The key element which differentiates this problem from standard high dimensional vector estimation lies in the structural assumptions which are formulated in this context. Indeed, the notion of sparsity assumption has been transposed into the concept of low-rank matrices and opened the way to numerous achievements (see for instance (Srebro, 2004; Cai et al., 2008)). In this paper, we argue that being low-rank is not only an equivalent of sparsity for matrices but that being low-rank and sparse can actually be seen as two orthogonal concepts. The underlying structure we have in mind is that of a block diagonal matrix. This situation occurs for instance in covariance matrix estimation in the case of groups of highly correlated variables or when denoising/clustering social graphs.

Efficient procedures developed in the context of sparse model estimation mostly rely on the use of $\ell_1$-norm regularization (Tibshirani, 1996). Natural extensions include cases where subsets of related variables are known to be active simultaneously (Yuan & Lin, 2006). These methods are readily adapted to matrix valued data and have been applied to covariance estimation (El Karoui, 2009; Bien & Tibshirani, 2010) and graphical model structure learning (Banerjee et al., 2007; Friedman et al., 2008). In the low-rank matrix completion problem, the standard relaxation approach leads to the use of the trace norm as the main regularizer within the optimization procedures (Srebro et al., 2005; Koltchinskii et al., 2011) and their resolution can either be obtained in closed form (loss measured in terms of Frobenius norm) or through iterative proximal solutions (Combettes & Pesquet, 2011; Beck & Teboulle, 2009) (for general classes of losses). However, solutions of low-rank estimation problems are in general not sparse at all, while denoising and variable selection on matrix-valued data are blind to the global structure of the matrix and process each variable independently.

In this paper, we study the benefits of using the sum of $\ell_1$ and trace-norms as regularizer. This sum of penalties on the same object allows to benefit from the virtues of both of them, in the same way as the elastic-net (Zou & Hastie, 2005) combines the sparsity-inducing property of the $\ell_1$ norm with the smoothness of the quadratic regularizer. Trace norm and $\ell_1$ penalties have already been combined in a different context.

In Robust PCA (Candes et al., 2009) and related literature, the signal $S$ is assumed to have an additive decomposition $S = X + Y$ where $X$ is sparse and $Y$ low-rank. Note that $S$ is not in general sparse nor low-rank and that this decomposition is subject to identifiability issues, as analyzed, e.g., in (Chandrasekaran et al., 2011). The decomposition is recovered by using $\ell_1$-norm regularization over $X$ and trace norm regularization over $Y$. This technique has been successfully applied to background substraction in image sequences, to graph clustering (Jalali et al., 2011) and covariance estimation (Luo, 2011).

Here, we consider the different situation where the matrix $S$ is sparse and low-rank at the same time. We demonstrate the applicability of our mixed penalty on different problems. We develop proximal methods to solve these convex optimization problems and we provide numerical evidence as well as theoretical arguments which illustrate the trade-off which can be achieved with the suggested method.

The remainder of the paper is organized as follows. In Section 2, we present the setup and motivations. Sections 3 and 4 are devoted to theoretical results on the interplay between sparse and low-rank effects. Section 5 presents algorithms used for resolution of the optimization problem and Section 6 is devoted to numerical experiments. The last Section explores related topics.

## 2. Setup and motivations

### 2.1. Problem formulation and notations

We first set some notations. For a matrix $S = (S_{i,j})_{i,j}$, we set the following matrix norms: $\|S\|_1 = \sum_{i,j} |S_{i,j}|$ and $\|S\|_* = \sum_{i=1}^{\text{rank}(S)} \sigma_i$, where $\sigma_i$ are the singular values of $S$ and $\text{rank}(S)$ is the rank of $S$. We consider the following setup. Let $A \in \mathbb{R}^{n \times n}$ be a fixed matrix and $\ell$ a loss function over matrices. We introduce the following optimization problem:

$$\arg\min_{S \in \mathcal{S}} \{\ell(S, A) + \gamma \|S\|_1 + \tau \|S\|_*\}$$

for some convex admissible set $\mathcal{S} \subset \mathbb{R}^{n \times n}$ and nonnegative regularization parameters $\gamma$, $\tau$.

In the sequel, the projection of a matrix $Z$ onto $\mathcal{S}$ is denoted by $P_{\mathcal{S}}(Z)$. The matrix $(M)_+$ is the componentwise positive part of the matrix M, and $\text{sgn}(M)$ is the sign matrix associated to $M$ with the convention $\text{sgn}(0) = 0$. The component wise product of matrices is denoted by $\circ$. The class $S_n^+$ of matrices is the convex cone of positive semidefinite matrices in $\mathbb{R}^{n \times n}$. The sparsity index of $M$ is $\|M\|_0 =$ $|\{M_{i,j} \neq 0\}|$ and the Frobenius norm of a matrix $M$ is defined by $\|M\|_F^2 = \sum_{i,j} M_{i,j}^2$. In Section 3, we shall also use $\|M\|_{\text{op}} = \sup_{x \,:\, \|x\|_2 = 1} \|Mx\|_2$ and $\|M\|_\infty = \max |M_{i,j}|$.

### 2.2. Main examples

The underlying assumption in this work is that the unknown matrix to be recovered has a block-diagonal structure. We now describe the main modeling choices through the following motivating examples:

- *Covariance matrix estimation* - the matrix $A$ represents a noisy estimate of the true covariance matrix obtained for instance with very few observations; the search space is $\mathcal{S} = S_n^+$ the class of positive semidefinite matrices; for the loss, we consider the squared norm $\ell(S, A) = \|S - A\|_F^2$.

- *Graph denoising* - the matrix $A$ is the adjacency matrix of a noisy graph with both irrelevant and missing edges; the search space is all of $\mathcal{S} = \mathbb{R}^{n \times n}$ and the coefficients of a candidate matrix estimate $S$ are interpreted as signed scores for adding/removing edges from the original matrix $A$; again, we use $\ell(S, A) = \|S - A\|_F^2$.

- *Link prediction* - the matrix $A$ is the adjacency matrix of a partially observed graph: entries are 0 for both not-existing and undiscovered links. The search space is unrestricted as before and the matrix $S$ contains the scores for link prediction; the ideal loss function is the empirical average of the zero-one loss for each coefficient

$$\ell_E(S, A) = \frac{1}{|E|} \sum_{(i,j) \in E} \mathbf{1}\{(A_{i,j} - 1/2) \cdot S_{i,j} \le 0\},$$

where $E$ is the set of edges in $A$. However, as in classification theory, practical algorithms should use a convex surrogate (e.g., the hinge loss).

## 3. Oracle inequality

The next result shows how matrix recovery is governed by the trade-off between the rank and the sparsity index of the unknown target matrix, or by their convex surrogates: the trace norm and the $\ell_1$-norm.

**Proposition 1.** *Let $S_0 \in \mathbb{R}^{n \times n}$ and $A = S_0 + \epsilon$ with $\epsilon \in \mathbb{R}^{n \times n}$ having i.i.d. entries with zero mean. Assume for some $\alpha \in [0; 1]$ that $\tau \ge 2\alpha \|\epsilon\|_{op}$ and $\gamma \ge 2(1 - \alpha) \|\epsilon\|_\infty$. Let*

$$\mathcal{L}(S) = \|S - A\|_F^2 + \tau \|S\|_* + \gamma \|S\|_1,$$

*and* $\widehat{S} = \arg\min_{S \in \mathcal{S}} \mathcal{L}(S)$ . *Then*

$$\|\widehat{S} - S_0\|_F^2 \leq \inf_{S \in \mathcal{S}} \left\{ \|S - S_0\|_F^2 + 2\tau \|S\|_* + 2\gamma \|S\|_1 \right\}$$

*and*

$$\|\widehat{S} - S_0\|_F^2 \leq \min \left\{ 2\tau \|S_0\|_* + 2\gamma \|S_0\|_1, \right.$$
$$\left. \left( \tau \sqrt{\mathrm{rank}(S_0)} \frac{\sqrt{2}+1}{2} + \gamma \sqrt{\|S_0\|_0} \right)^2 \right\}.$$

The techniques used in the proof (see the Appendix) are very similar to those introduced in (Koltchinskii et al., 2011). Note that the upper bound interpolates between the results known for trace-norm penalization and Lasso. In fact, for $\alpha = 0$, $\tau$ can be set to zero, and we get a sharp bound for Lasso, while the trace-norm regression bounds of (Koltchinskii et al., 2011) are obtained for $\alpha = 1$.

## 4. Generalization error in link prediction

We dwell for a moment on the task of link prediction in order to illustrate how rank and sparsity constraints can help in this setting. Given a subset E of observed edges from a graph adjacency matrix $A \in \{0,1\}^{n \times n}$, we set out to predict unobserved links by finding a sparse rank $r$ predictor $S \in \mathbb{R}^{n \times n}$ with small zero-one loss

$$\ell(S, A) = \frac{1}{n^2} \sum_{(i,j) \in \{1,\ldots,n\}^2} \mathbb{1}\{(A_{i,j} - 1/2) \cdot S_{i,j} \leq 0\}$$

by minimizing the empirical zero-one loss

$$\ell_E(S, A) = \frac{1}{|E|} \sum_{(i,j) \in E} \mathbb{1}\{(A_{i,j} - 1/2) \cdot S_{i,j} \leq 0\}.$$

The objective of a generalization bound is to relate $\ell(S, A)$ with $\ell_E(S, A)$. In the case of the sole rank constraint, (Srebro, 2004) remarked that all low-rank matrices with the same sign pattern are equivalent in terms of loss and applied a standard argument for generalization in classes of finite cardinality. In the work of Srebro, a beautiful argument is used to upper bound the number of distinct sign configurations for predictors of rank $r$

$$s_{\mathrm{lr}}(n, r) = |\{\mathrm{sgn}(S) \mid S \in \mathbb{R}^{n \times n}, \mathrm{rank}(S) \leq r\}|$$

leading to the following generalization performance: for $\delta > 0$, $A \in \{0,1\}^{n \times n}$ and with probability $1 - \delta$ over choosing a subset E of entries in $\{1, \ldots, n\}^2$ uniformly among all subsets of $|E|$ entries, we have for any matrix $S$ of rank at most $r$ and $\Delta(n, r) = \left( \frac{8en}{r} \right)^{2nr}$

$$\ell(S, A) < \ell_E(S, A) + \sqrt{\frac{\log \Delta(n, r) - \log \delta}{2|E|}}. \quad (1)$$

We consider the class of sparse rank $r$ predictors

$$\mathcal{M}(n, r, s) = \{UV^T \mid U, V \in \mathbb{R}^{n \times r}, \|U\|_0 + \|V\|_0 \leq s\}$$

and let $s_{\mathrm{splr}}(n, r, s)$ be the number of sign configurations for the set $\mathcal{M}(n, r, s)$. By upper bounding the number of sign configurations for a fixed sparsity pattern in $(U, V)$ using an argument similar to (Srebro, 2004), a union bound gives

$$s_{\mathrm{splr}}(n, r, s) \leq \Gamma(n, r, s) = \left( \frac{16en^2}{s} \right)^s \binom{2nr}{s}.$$

Using the same notations as previously, we deduce from this result the following generalization bound: with probability $1 - \delta$ and for all $S \in \mathcal{M}(n, r, s)$,

$$\ell(S, A) < \ell_E(S, A) + \sqrt{\frac{\log \Gamma(n, r, s) - \log \delta}{2|E|}}. \quad (2)$$

In general, bound (2) is tighter than (1) for sufficiently large values of $n$ as shown in the next proposition. The two bounds coincide when $s = 2nr$, that is, when $(U, V)$ is dense and there is no sparsity constraint.

**Proposition 2.** *For $r_n = n\beta$ with $\beta \in ]0,1]$ and $s_n = n\alpha$ with $\alpha \leq 2\beta$,*

$$\frac{\Delta(n, r_n)}{\Gamma(n, r_n, s_n)} = \Omega \left( \left[ \frac{8en(\beta n - \alpha)}{(\beta n)^2} \right]^{2n^2\beta} \right),$$

*which diverges when $n$ goes to infinity.*

*Proof.* The result follows from the application of Stirling's formula. $\square$

By considering a predictor class of lower complexity than low-rank matrices, we can thus achieve better generalization performances.

## 5. Algorithms

We now present how to solve the optimization problem with mixed penalties presented in Section 2. We consider a loss function $\ell(S, A)$ convex and differentiable in $S$, and assume that its gradient is Lipschitz with constant $L$ and can be efficiently computed. This is, in particular, the case for the squared Frobenius norm previously mentioned and for other classical choices such as the hinge loss.

## 5.1. Proximal operators

We encode the presence of a constraint set $\mathcal{S}$ using the indicator function $1_{\mathcal{S}}(S)$ that is zero when $S \in \mathcal{S}$ and $+\infty$ otherwise, leading to

$$\hat{S} = \underset{S \in \mathbb{R}^{n \times n}}{\arg \min} \left\{ \ell(S, A) + \gamma ||S||_* + \tau ||S||_1 + 1_{\mathcal{S}}(S) \right\}.$$

This formulation involves a sum of a convex differentiable loss and of convex non differentiable regularizers which renders the problem non trivial. A string of algorithms have been developed for the case where the optimal solution is easy to compute when each regularizer is considered in isolation. Formally, this corresponds to cases where the proximal operator defined for a convex regularizer $R : \mathbb{R}^{n \times n} \to \mathbb{R}$ at a point $Z$ by

$$\text{prox}_R(Z) = \underset{S \in \mathbb{R}^{n \times n}}{\arg \min} \frac{1}{2} ||S - Z||_F^2 + R(S).$$

is easy to compute for each regularizer taken separately. See (Combettes & Pesquet, 2011) for a broad overview of proximal methods.

The proximal operator of the indicator function is simply the projection onto $\mathcal{S}$, which justifies the alternate denomination of generalized projection operator for $\text{prox}_R$. The proximal operator for the trace norm is given by the shrinkage operation as follows (Beck & Teboulle, 2009). If $Z = U \text{diag}(\sigma_1, \cdots, \sigma_n) V^T$ is the singular value decomposition of $Z$,

$$\text{SHR}_\tau(Z) := \text{prox}_{\tau ||.||_*}(Z) = U \text{diag}((\sigma_i - \tau)_+)_i V^T.$$

Similarly, the proximal operator for the $\ell_1$-norm is the soft thresholding operator

$$\text{ST}_\gamma(Z) := \text{prox}_{\gamma ||.||_1} = \text{sgn}(Z) \circ (|Z| - \gamma)_+.$$

## 5.2. Generalized Forward-Backward splitting

The family of Forward-Backward splitting methods are iterative algorithms applicable when there is only one non differentiable regularizer. These methods alternate a gradient step and and a proximal step, leading to updates of the form

$$S_{k+1} = \text{prox}_{\theta R}(S_k - \theta \, \text{grad}_S \, \ell(S, A)).$$

In particular, this corresponds to projected gradient descent when $R$ is the indicator function of a convex set. On the other hand, Douglas-Rachford splitting tackles the case of $q \geq 2$ terms but does not benefits from differentiability. A generalization of these two setups has been recently proposed in (Raguet et al., 2011) under the name of Generalized Forward-Backward, which we specialize to our problem in Algorithm 1. The proximal operators are applied in parallel, and the resulting $(Z_1, Z_2, Z_3)$ is projected onto the constraint that $Z_1 = Z_2 = Z_3$ which is given by the mean. The auxiliary variable $Z_3$ can be simply dropped when $\mathcal{S} = \mathbb{R}^{n \times n}$. The algorithm converges under very mild conditions when the step size $\theta$ is smaller than $\frac{2}{L}$.

---

**Algorithm 1** Generalized Forward-Backward

Initialize $S, Z_1, Z_2, Z_3 = A$, q = 3
**repeat**
  Compute $G = \nabla_S \ell(S, A)$.
  Compute $Z_1 = \text{prox}_{q \theta \tau ||.||_*}(2S - Z_1 - \theta G)$
  Compute $Z_2 = \text{prox}_{q \theta \gamma ||.||_1}(2S - Z_2 - \theta G)$
  Compute $Z_3 = P_{\mathcal{S}}(2S - Z_3 - \theta G)$
  Set $S = \frac{1}{q} \sum_{k=1}^{q} Z_k$
**until** convergence
**return** S

---

## 5.3. Incremental Proximal Descent

Although Algorithm 1 performs well in practice, the $O(n^2)$ memory footprint with a large leading constant due to the parallel updates can be a drawback in some cases. As a consequence, we mention a matching serial algorithm (Algorithm 2) introduced in (Bertsekas, 2011) that has a flavor similar to multi-pass stochastic gradient descent. We present here a version where updates are performed according to a cyclic order, although random selection of the order of the updates is also possible.

---

**Algorithm 2** Incremental Proximal Descent

Initialize $S = A$
**repeat**
  Set $S = S - \theta \nabla_S \ell(S, A)$
  Set $S = \text{prox}_{\theta \tau ||.||_*}(S)$
  Set $S = \text{prox}_{\theta \gamma ||.||_1}(S)$
  Set $S = P_{\mathcal{S}}(S)$
**until** convergence
**return** S

---

## 5.4. PSD constraint

For any positive semidefinite matrix, we have $||Z||_* = \text{Tr}(Z)$. The simple form of the trace norm allows to take into account the positive semidefinite constraint at no additional cost, as the shrinkage operation and the projection onto the convex cone of positive semidefinite matrices can be combined into a single operation.

**Lemma 1.** *For* $\tau \geq 0$ *and* $S \in \mathbb{R}^{n \times n}$,

$$\text{prox}_{\tau \|.\|_* + 1_{S_n^+}}(S) = \underset{Z \succeq 0}{\arg\min} \frac{1}{2} \|Z - S\|_F^2 + \tau \|Z\|_*$$
$$= P_{S_n^+}(S - \tau I_n).$$

## 6. Numerical experiments

We present numerical experiments to highlight the benefits of our method. For efficiency reasons, we use the serial proximal descent algorithm (Algorithm 2).

### 6.1. Synthetic data

*Covariance matrix estimation.* We draw $N$ vectors $x_i \sim \mathcal{N}(0, \Sigma)$ for a block diagonal covariance matrix $\Sigma \in \mathbb{R}^{n \times n}$. We use $r$ blocks of random sizes and of the form $vv^\top$ where the entries of $v$ are drawn i.i.d. from the uniform distribution on $[-1, 1]$. Finally, we add gaussian noise $\mathcal{N}(0, \sigma^2)$ on each entry. In our experiments $r = 5$, $N = 20$, $n = 100$, $\sigma = 0.6$. We apply our method (SPLR), as well as trace norm regularization (LR) and $\ell_1$ norm regularization (SP) to the empirical covariance matrix, and report average results over ten runs. Figure 1 shows the RMSE normalized by the norm of $\Sigma$ for different values of $\tau$ and $\gamma$. Note that the effect of the mixed penalty is visible as the minimum RMSE is reached inside the $(\tau, \gamma)$ region. We perform, on the same data, separate cross-validations on $(\tau, \gamma)$ for SPLR, on $\tau$ for LR and on $\gamma$ for SP. We show in Figure 2 the supports recovered by each algorithm, the output matrix of LR being thresholded in absolute value. The support recovery demonstrates how our approach discovers the underlying patterns despite the noise and the small number of observations.

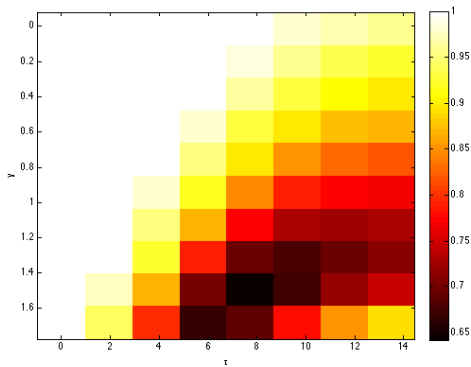

*Figure 1.* Covariance estimation. Cross-validation: normalized RMSE scores (SPLR)

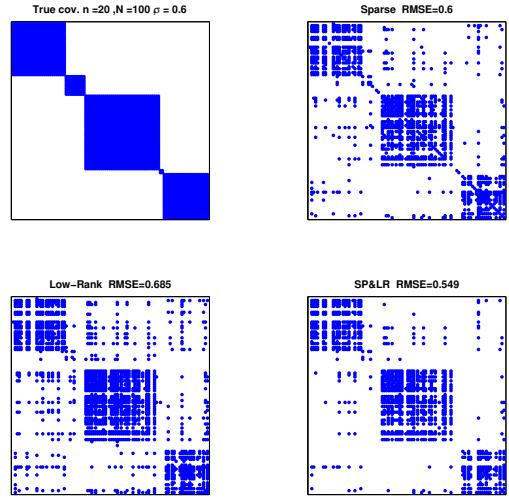

*Figure 2.* Covariance estimation. Support of $\Sigma$ (top left), and of the estimates given by SP (top right), LR (bottom left), and SPLR (bottom right)

### 6.2. Real data sets

*Protein Interactions.* We use data from (Hu et al., 2009), in which protein interactions in Escherichia coli bacteria are scored by strength in $[0, 2]$. The data is, by nature, sparse. In addition to this, it is often suggested that interactions between two proteins are governed by a small set of factors, such as surface accessible amino acid side chains (Bock & Gough, 2001), which motivates the estimation of a low-rank representation. Representing the data as a weighted graph, we filter to retain only the 10% of all 4394 proteins that exhibit the most interactions as measured by weighted degree. We corrupt 10% of entries of the adjacency matrix selected uniformly at random by uniform noise in $[0, \eta]$. Parameters are selected by cross-validation and algorithms are evaluated using mean RMSE between estimated and original adjacency matrices over 25 runs. RMSE scores are shown in Table 1 and show the empirical superiority of our approach (SPLR).

| $\eta$ | SPLR | LR | SP |
|---|---|---|---|
| 0.1 | **0.0854** ±0.012 | 0.1487 ±0.02 | 0.1023 ±0.02 |
| 0.2 | **0.2073** ± 0.03 | 0.2673 ± 0.3 | 0.2484 ± 0.03 |
| 0.3 | 0.3105 ± 0.03 | 0.3728 ± 0.03 | **0.3104** ± 0.02 |

*Table 1.* Prediction of interactions in Escherichia coli. Mean normalized RMSE and standard deviations.

*Social Networks.* We have performed experiments with the Facebook100 data set analyzed by (Traud et al., 2011). The data set comprises all friendship relations between students affiliated to a specific university, for a selection of one hundred universities. We select a single university with 41554 users and filter as in the previous case to keep only the 10% users with highest degrees. In this case, entries are corrupted by impulse noise: a fixed fraction $\sigma$ of randomly chosen edges are flipped, thus introducing noisy friendship relations and masking some existing relations. The task is to discover the noisy relations and recover masked relations. We compare our method to standard baselines in link prediction (Liben-Nowell & Kleinberg, 2007). Nearest Neighbors (NN) relies on the number of common friends between each pair of users, which is given by $A^2$ when $A$ is the noisy graph adjacency matrix. Katz's coefficient connects a pair of nodes according to a score based on the number of paths connecting them, emphasizing short paths. Results are reported in Table 2 using the area under the ROC curve (AUC). SPLR outperforms LR but also NN and Katz which do not directly seek a low-rank representation.

| $\sigma$ | SPLR | LR | NN | Katz |
|------|--------|--------|--------|--------|
| 5 % | 0.9293 | 0.9291 | 0.7680 | **0.9298** |
| 10 % | **0.9221** | 0.9174 | 0.7620 | 0.9189 |
| 15 % | **0.9117** | 0.9024 | 0.7555 | 0.9068 |
| 20 % | **0.8997** | 0.8853 | 0.7482 | 0.8941 |

*Table 2.* Facebook denoising data. Mean AUC over 10 simulation runs. All standard deviations are lower than $3 \cdot 10^{-4}$.

## 7. Discussion

*Other loss functions.* The methods presented in this paper can be seamlessly extended to non-square matrices, which can arise, for instance, from adjacency matrices of bipartite graphs. Our work also applies to a wide range of other losses. A useful example that links our work to the matrix completion framework is when linear measurements of the target matrix or graph are available, or can be predicted as in (Richard et al., 2010). In this case, the loss can be defined in the feature space. Due to the low-rank assumption, our method does not directly apply to the estimation of precision matrices often used for gaussian graphical model structure learning (Friedman et al., 2008), and the applications of conditional independence structures generated by low-rank and possibly sparse models is to be discussed. Note that the trace norm constraint is vacuous for some special classes of positive semi-definite matrices. For instance, it is not useful for estimating a correlation matrix as, in this case, the trace is always equal to the dimension.

*Matrix factorizations.* A related and popular task is finding low-rank factorizations of matrices of the form

$UV^T$ (see, e.g., (Srebro et al., 2005; Srebro, 2004)), thus jointly optimizing in $U, V \in \mathbb{R}^{n \times r}$ loss functions of the form $\ell((U, V), A) = ||UV^T - A||_F^2$ for some target maximum rank $r$. This implicitly encodes the low-rank constraint which leads to efficient optimization schemes, and allows for interpretability as estimated $(U, V)$ pairs can be considered as latent factors. Non-negative Matrix Factorization (NMF) (Lee et al., 1999) imposes non negativity constraints on the coefficients of U and V to enhance interpretability by allowing only for additive effects and tends to produce sparse factor matrices $U, V$, although this a rather indirect effect. There is no strong guarantee on the sparsity achieved by NMF nor is it easy to set the target sparsity and different methods for sparse NMF have been proposed in (Hoyer, 2004; Kim & Park, 2008). Sparse matrix factorizations have also been proposed without the positivity constraint. Most work along this line is motivated by extending the classical PCA and finding sparse directions that maximize the variance of the projection. Most methods give up orthogonality between the components and can thus be seen as sparse matrix factorization techniques. SPCA proposed in (Zou et al., 2004) penalizes the $\ell_1$ norm of the principal components and can be reduced to solving independent elastic-nets. A different formulation using SDP programming is introduced in (D'Aspremont et al., 2007) with good empirical results. In spite of good empirical performances, all these methods based on matrix factorization suffer from a significant drawback. Although formulations are usually convex in U or V, they are not in general jointly convex and optimization procedures can get stuck in local minima.

*Regularization parameters.* We showed how to empirically select using cross-validation the hyper parameters $\tau$ and $\gamma$ for a specific application. From a theoretical point of view, Proposition 1 provides us with performance guarantees when the regularization parameters are large enough. We know from random matrix theory that the operator norm of a random gaussian matrix concentrates around $\sqrt{n}$ which enforces a stringent constraint on $\tau$ for $\tau \geq 2\alpha||\epsilon||_{op}$ to hold with high probability. Similarly, the $\infty$-norm $||\epsilon||_\infty$ can be bounded by $||\epsilon||_{op}$ or using the multivariate Tchebycheff inequality of (Olkin & Pratt, 1958) which implies that the condition $\gamma \geq 2(1 - \alpha)||\epsilon||_\infty$ is satisfied with probability $1 - \delta$ when $\gamma = \Omega\left((1 - \alpha)\frac{2n\sigma}{\delta}\right)$. In practice, $\gamma$ should not exceed the order of magnitude of the entries of the matrix, as this leads to a trivial zero solution. Asymptotically, to keep the sparsity regularization parameter $\gamma$ of the order of magnitude of elements of the observation matrix $A$, the free parameter $\alpha$ must be chosen so that $1 - \alpha_n \sim_n \frac{1}{n}$. This gives the same asymptotic behavior in $O(\sqrt{n})$ for the lower bound on $\tau$ as in matrix completion.

*Optimization.* Other optimization techniques can be considered for future work. A trace norm constraint alone can be taken into account without projection or relaxation into a penalized form by casting the problem as a SDP as proposed in (Jaggi, 2011). The special form of this SDP can be leveraged to use the efficient resolution technique from (Hazan, 2008). This method applies to a differentiable objective whose curvature determines the performances. Extending these methods with projection onto the $\ell_1$ ball or a sparsity-inducing penalty could lead to interesting developments.

## Appendix- Sketch of proof for Prop. 1

For any S in $\mathcal{S}$ and by optimality of $\widehat{S}$,

$$-2\langle \widehat{S} - S, S_0\rangle \leq -2\langle \widehat{S} - S, S_0\rangle + \mathcal{L}(S) - \mathcal{L}(\widehat{S})$$

$$\leq 2\alpha\|\widehat{S} - S\|_*\|\epsilon\|_{op} + 2(1-\alpha)\|\widehat{S} - S\|_1\|\epsilon\|_\infty$$

$$+ \tau(\|S\|_* - \|\hat{S}\|_*) + \gamma(\|S\|_1 - \|\widehat{S}\|_1) + \|S\|_F^2 - \|\hat{S}\|_F^2$$

for any $\alpha \in [0; 1]$. The assumptions on $\tau, \gamma$ and triangular inequality lead to the first bound.

Let $r = \text{rank}(S)$, $k = \|S\|_0$, $S = \sum_{j=1}^r \sigma_j u_j v_j^\top$ the SVD of $S$, $S = \Theta \circ |S|$, where $\Theta = \text{sgn}(S)$, and $\Theta^\perp \in \{0,1\}^{n\times n}$ the complementary sparsity pattern. We use $P_{S_1^\perp}$ (resp. $P_{S_2^\perp}$) to denote the projection operator onto the orthogonal of the left (resp. right) singular space of $S$. We also note $\mathcal{P}_S(X) = X - P_{S_1^\perp} X \mathcal{P}_{S_2^\perp}$ such that $X = \mathcal{P}_S(X) + P_{S_1^\perp} X \mathcal{P}_{S_2^\perp}$.

Any element $V$ of the subgradient of the convex function $S \mapsto \tau\|S\|_* + \gamma\|S\|_1$ can be decomposed as

$$V = \tau\left(\sum_{j=1}^r u_j v_j^\top + P_{S_1\perp} W_* P_{S_2\perp}\right) + \gamma\left(\Theta + W_1 \circ \Theta^\perp\right)$$

for $W_1, W_*$ with $\|W_*\|_{op} \leq 1$, $\|W_1\|_\infty \leq 1$, which can be chosen such that

$$\langle V, \widehat{S} - S\rangle = \tau\langle\sum_{j=1}^r u_j v_j^\top, \widehat{S} - S\rangle + \tau\|P_{S_1\perp}\widehat{S}P_{S_2\perp}\|_*$$

$$+ \gamma\langle\Theta, \widehat{S} - S\rangle + \gamma\|\Theta^\perp \circ \widehat{S}\|_1.$$

By monotonicity of the subdifferential and optimality conditions,

$$2\langle\widehat{S} - S_0, \widehat{S} - S\rangle$$

$$\leq 2\langle\epsilon, \widehat{S} - S\rangle - \tau\langle\sum_{j=1}^r u_j v_j^\top, \widehat{S} - S\rangle$$

$$- \tau\|P_{S_1^\perp}\widehat{S}P_{S_2^\perp}\|_* - \gamma\langle\Theta, \widehat{S} - S\rangle - \gamma\|\Theta^\perp \circ \widehat{S}\|_1.$$

Decompose

$$\epsilon = \alpha\left(\mathcal{P}_S(\epsilon) + P_{S_1^\perp}\epsilon P_{S_2^\perp}\right) + (1-\alpha)\left(|\Theta| \circ \epsilon + \Theta^\perp \circ \epsilon\right).$$

Using results on dual norms, we have

$$|\langle M_1, M_2\rangle| \leq \|M_1\|_*\|M_2\|_{op}$$

$$|\langle M_1, M_2\rangle| \leq \|M_1\|_1\|M_2\|_\infty$$

for all $M_1, M_2 \in \mathbb{R}^{n\times n}$ and hence,

$$\langle\epsilon, \widehat{S} - S\rangle \leq \alpha\|\mathcal{P}_S(\epsilon)\|_F\|P_{S_1}(\widehat{S} - S)P_{S_2}\|_F$$

$$+ \alpha\|P_{S_1^\perp}\epsilon P_{S_2^\perp}\|_{op}\|P_{S_1^\perp}\widehat{S}P_{S_2^\perp}\|_*$$

$$+ (1-\alpha)\|\Theta \circ \epsilon\|_F\|\Theta \circ (\widehat{S} - S)\|_F$$

$$+ (1-\alpha)\|\Theta^\perp \circ \epsilon\|_\infty\|\Theta^\perp \circ \widehat{S}\|_1.$$

Using

$$\|\mathcal{P}_S(\epsilon)\|_F \leq \sqrt{2\,r}\|\epsilon\|_{op}, \quad \|\Theta \circ \epsilon\|_F \leq \sqrt{k}\|\epsilon\|_\infty$$

leads for $\tau \geq 2\alpha\|\epsilon\|_{op}$ and $\gamma \geq 2(1-\alpha)\|\epsilon\|_\infty$ to

$$\|\widehat{S} - S_0\|_F^2 + \|\widehat{S} - S\|_F^2$$

$$\leq \|S - S_0\|_F^2 + \left(\tau\sqrt{r}(\sqrt{2}+1) + 2\gamma\sqrt{k}\right)\|\widehat{S} - S\|_F.$$

Using $\beta x - x^2 \leq \left(\frac{\beta}{2}\right)^2$, we obtain

$$\|\widehat{S} - S_0\|_F^2 \leq \|S - S_0\|_F^2 + \frac{1}{4}\left(\sqrt{r}\tau(\sqrt{2}+1) + 2\sqrt{k}\gamma\right)^2$$

and setting $S = S_0$ gives the result. $\square$